\documentclass[%
 reprint,
 amsmath,amssymb,amsfont
 prl,
]{revtex4-2}

\usepackage[utf8]{inputenc}
\usepackage[english]{babel}
\usepackage{graphicx}
\usepackage{dcolumn}
\usepackage{bm}
\usepackage{comment}
\usepackage[table,dvipsnames]{xcolor}
\usepackage{tikz}
\usepackage{subcaption}
\usepackage{siunitx}
\usepackage{booktabs}
\usepackage{xr-hyper} 
\usepackage{hyperref}

\makeatletter
\newcommand*{\addFileDependency}[1]{
  \typeout{(#1)}
  \@addtofilelist{#1}
  \IfFileExists{#1}{}{\typeout{No file #1.}}
}
\makeatother

\newcommand*{\myexternaldocument}[1]{%
    \externaldocument{#1}%
    \addFileDependency{#1.tex}%
    \addFileDependency{#1.aux}%
}
\myexternaldocument{appendix}

\usepackage{multirow}

\newcommand{\dd}{\partial}

\newcommand{\uu}{\mathbf{u}}
\newcommand{\UU}{\mathbf{U}}

\newcommand{\WW}{\mathbf{W}}

\newcommand{\bF}{\mathbf{F}}

\newcommand{\bP}{\mathbf{P}}

\newcommand{\bsub}{\begin{subequations}}
\newcommand{\esub}{\end{subequations}}
\newcommand{\ba} {\begin{align} }
\newcommand{\ea}{ \end{align} }

\begin{document}

\UseRawInputEncoding 


\title{Laws of mutual spiral wave interaction in excitable media}

\author{Tim De Coster\textsuperscript{1,5},Arstanbek Okenov\textsuperscript{1}, Debora Hoogendijk\textsuperscript{2},
Arman Nobacht\textsuperscript{1}, Mathilde Rivaud\textsuperscript{1}, Antoine A.F. de Vries\textsuperscript{1}, 
, Dani\"el Pijnappels\textsuperscript{1}, Vivi Rottsch\"afer\textsuperscript{2,3}, Hans Dierckx\textsuperscript{1,2,4} }
 \email{h.j.f.dierckx@lumc.nl}
\affiliation{
 \textsuperscript{1} Laboratory of Experimental Cardiology, Leiden University Medical Center (LUMC), Leiden, The Netherlands, 
 \textsuperscript{2} Mathematical Institute, Leiden University, Leiden, the Netherlands
 \textsuperscript{3} Korteweg-de Vries Institute for Mathematics, University of Amsterdam, 
Amsterdam, The Netherlands
 \textsuperscript{4} Leiden Institute of Physics (LION), Leiden University, Leiden, the Netherlands
 \textsuperscript{5} Max Planck Institute for Dynamics and Self-Organisation, G\"{o}ttingen, Germany
}
\date{\today}



\begin{abstract}
Interacting rotating spiral waves have been observed in complex systems, such as cardiac fibrillation, cognitive processing in the brain cortex and oscillating chemical reactions, during dynamical regimes that are still poorly understood. 
We present the equivalent of Newton's gravitational attraction law for spiral waves on planar reaction-diffusion systems. 
The spiral waves' phases and positions determine their regions of influence, separated by collision interfaces. At the collision interfaces, wave front deflections cause spiral drift that pushes the interfaces forward. As a result, the spiral wave drift velocity is proportional to the total force exerted on on it, which can be determined by a boundary integral over its region of influence. The proportionality factor between force and response is akin to the `mass' of the spiral. However, this spiral mass depends on the region of influence of the spiral and thus also varies over time. The forces between spiral wave pairs are not directed along the line connecting their centers, violating Newton's law of action and reaction. Our solution to the N-body interaction problem for spirals in extended excitable media encompasses both pairwise interactions and spiral wave drift in bounded domains, with application to cardiac fibrillation. 
\end{abstract}

\maketitle

\section{Introduction.} Rotating spiral waves are robust patterns that originate in a wide variety of physical, chemical and biological systems\cite{Zhabotinsky:1973,Allesie:1976,Rotermund:1990,xu_interacting_2023}. Due to their topological charge, spiral waves are created in pairs, can persist for long times in the system and therefore often determine the system's appearance or function. Prime examples are rotating vortices in the Belaousov-Zhabotinksy chemical reaction \cite{Zhabotinsky:1973} and electrical rotors in the heart during cardiac arrhythmias \cite{Allessie:1973,Gray:1995}. In the cardiac context, a fast heart rate (tachycardia) is typically associated with a single spiral wave, while during life-threatening fibrillation in either the atria or ventricles multiple interacting spirals are observed \cite{Gray:1995}. Specifically for spiral interactions, mutual attraction between nearby spirals of opposite rotation sense leads to their annihilation, thereby favoring simpler patterns. 

Although interacting spiral waves emerged even from the first numerical model of atrial fibrillation \cite{Moe:1964}, the laws of mutual spiral interaction have not yet been revealed. Previous studies on the evolution of single spirals under external influences such as parameter gradients \cite{Vinson:1999, Henry:2004}, feedback stimulation \cite{Biktashev:1995b}, domain curvature \cite{Davydov:2000} and anisotropic wave speeds \cite{Dierckx:2013} have shown that they respond in a clear, predictably manner to stimuli, thereby opening avenues to control them. A key observation is that spiral waves have particle-like properties, also called particle-wave duality \cite{Biktasheva:2003}. This statement follows from the exponential localization of the spiral wave's sensitivity to perturbations, as described by its adjoint critical eigenfunctions, also called response functions \cite{Keener:1986, Barkley:1992, Biktasheva:1998}. 

\begin{figure*}[t]
    \centering
    \raisebox{4cm}{a)}
    \includegraphics[width=0.25\linewidth]{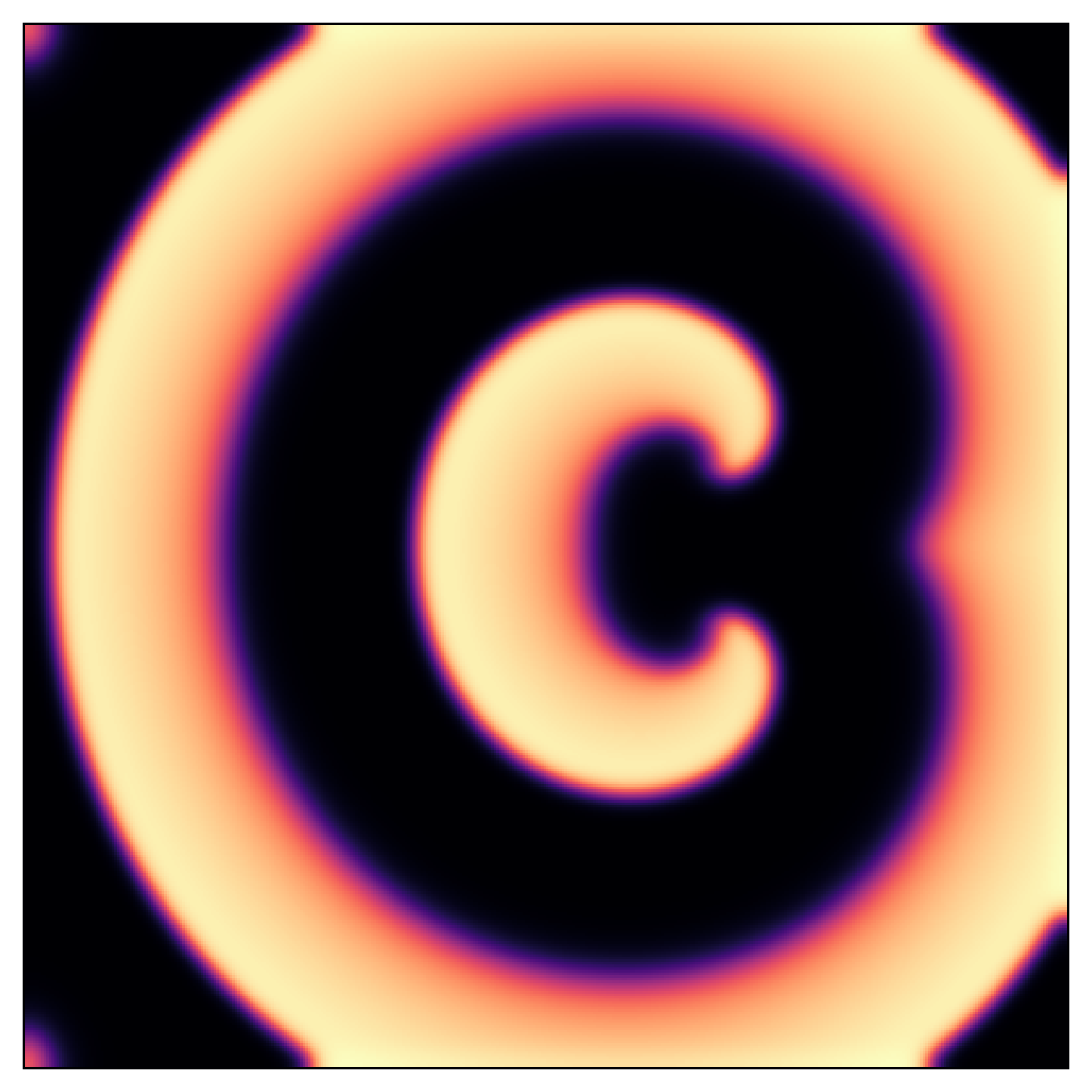}
\raisebox{4cm}{b)}
    \includegraphics[width=0.25\linewidth]{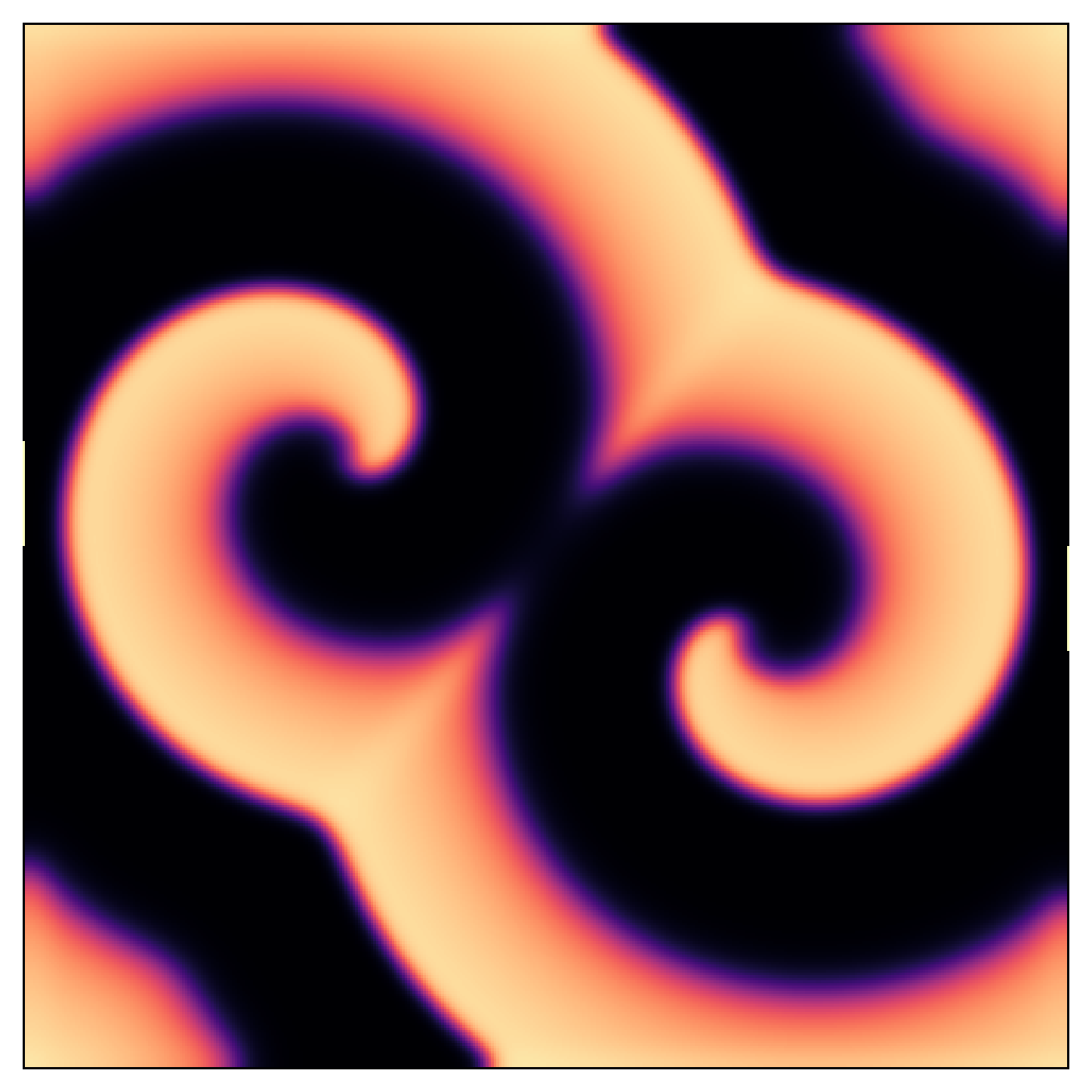}
\raisebox{4cm}{c)}
\includegraphics[width=0.25\linewidth]{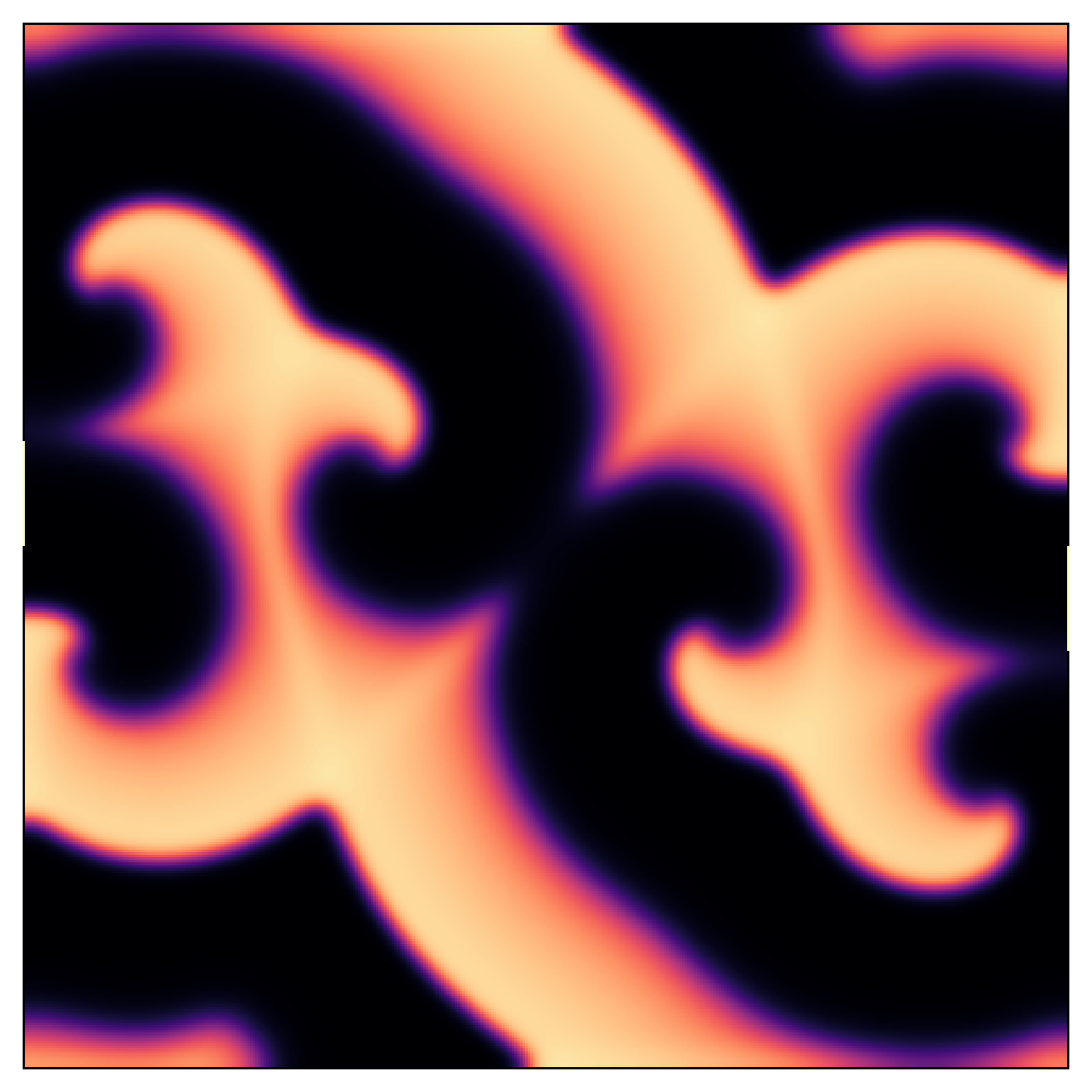} \\
\raisebox{4cm}{d)}
    \includegraphics[width=0.25\linewidth]{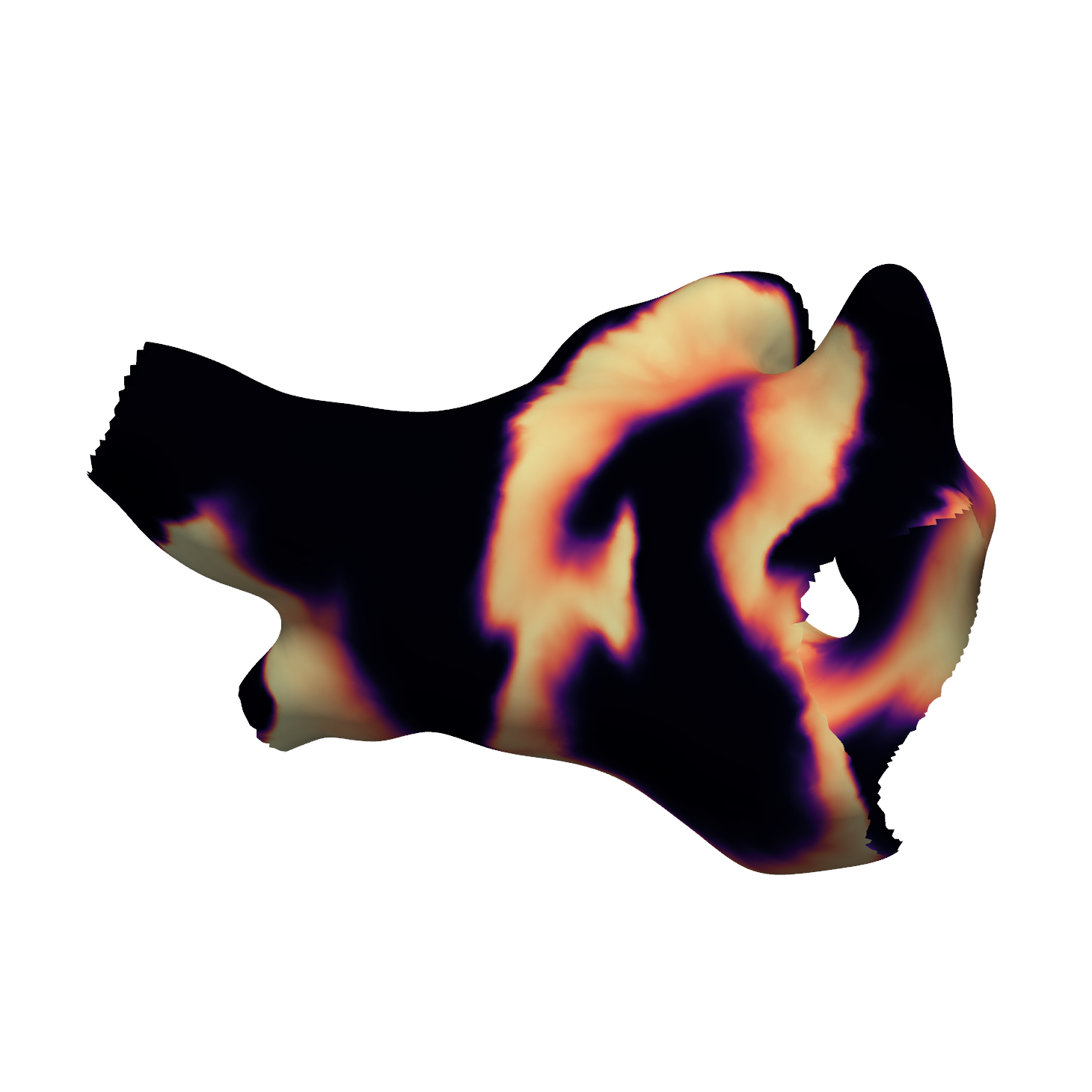}
\raisebox{4cm}{e)}
    \includegraphics[width=0.25\linewidth]{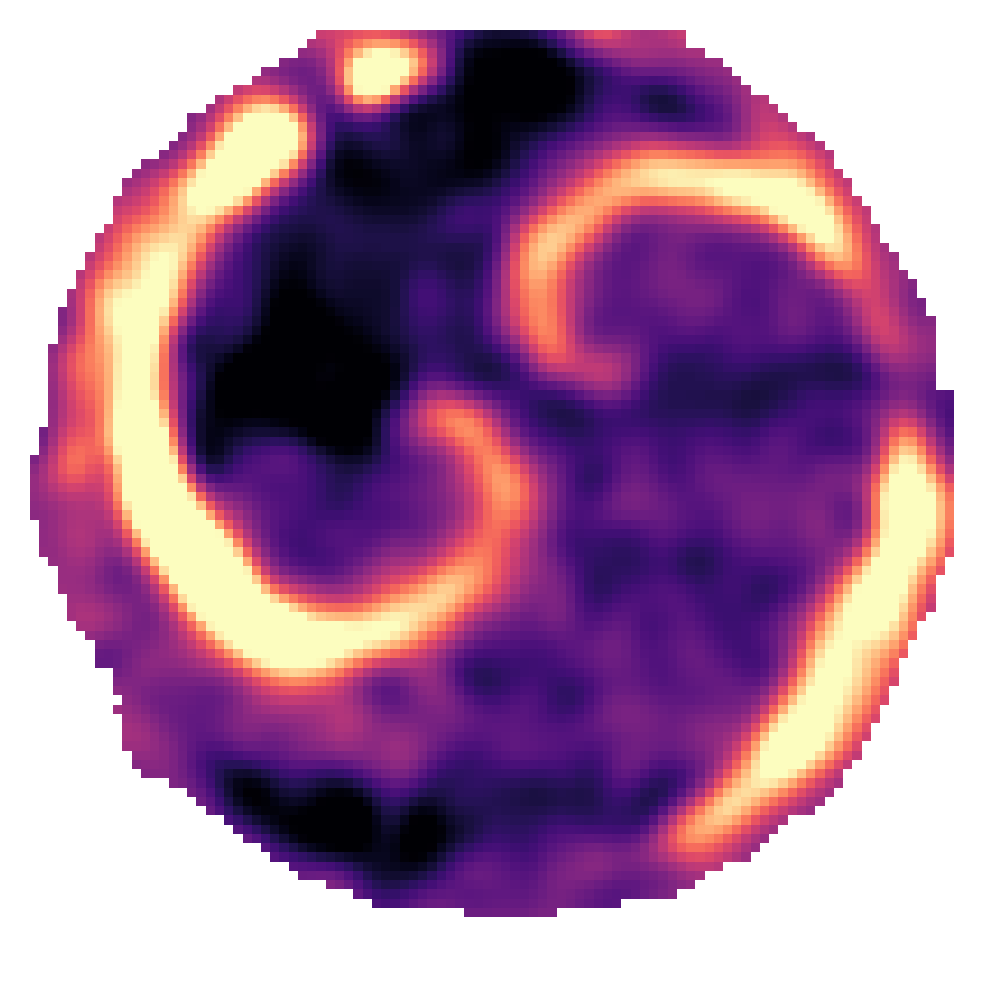}
\raisebox{4cm}{f)}
    \includegraphics[width=0.25\linewidth]{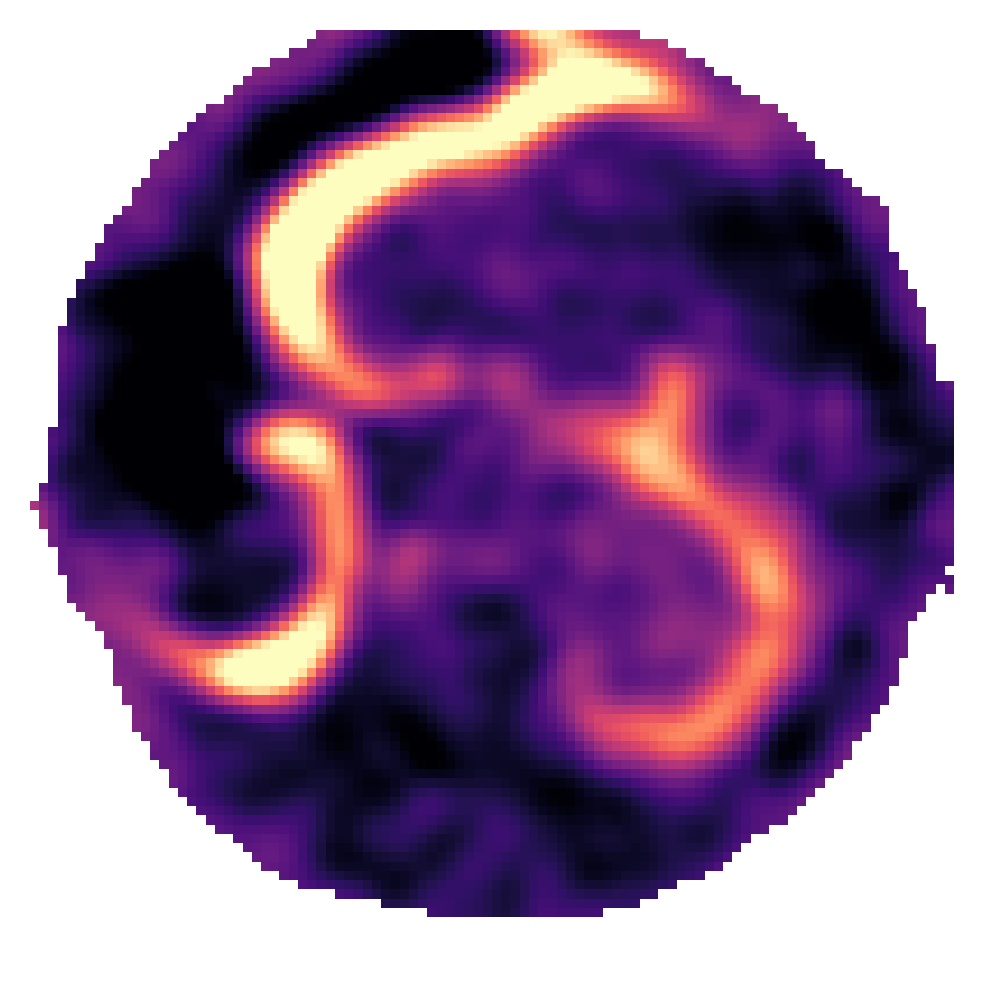}
    \caption{Examples of multiple interacting spirals in excitable media, in simulations \cite{Aliev:1996} (a-d) and \textit{in vitro} experiments \cite{Harlaar:2021} (e-f). (a) Pair of oppositely rotating spirals, equivalent to a single spiral near a planar boundary. (b) Spiral pair with the same rotation sense. (c) Multiple spiral simulation. (d) Spiral wave simulation on a curved surface with the geometry of the atria of a human heart. (e) Culture of conditionally immortalized human atrial myocytes showing a spiral pair. (f) Same set-up, with multiple interacting spirals.
    \label{fig:intro}}
\end{figure*}

Numerical simulation of multiple spiral waves demonstrated that they may tend to cluster together in pairs or triplets \cite{Zaritski:2002} and also form asymmetric bound pairs \cite{zemlin_asymmetric_2005}. 

The problem of spiral wave interaction is closely related to spiral-boundary interaction, as illustrated by the method of mirrors \cite{Jackson:1975}. If a single spiral wave has its rotation center at a distance $d$ from a no-flux boundary, mirroring it across the boundary produces a pair of synchronized spirals rotating in opposite directions, separated by a distance of $2d$. Spiral-boundary interaction has been extensively studied, as the boundary provides the only location where a single spiral wave can be removed from the medium. Therefore, boundary interaction is key to the successful paced control of spiral waves. If tailored stimulation drives a spiral wave toward the boundary but it stabilizes there before being eliminated, the control method fails. In the cardiac atria, the topology includes several openings where veins attach. Although it has long been known that only openings with a perimeter larger than the wavelength can sustain anatomical reentry, a comprehensive theory is still lacking for predicting when such `holes' in the medium will attract spiral waves, stabilize them, and determine the resulting dynamical regime. 

The problems described above can be viewed as an analogue of the N$‑$body problem for spiral waves, echoing the historical effort to understand the collective dynamics of planetary motion. 
This work provides answers to these questions via mathematical analysis of the evolution equations that govern a wide class of continuous excitable media, namely the reaction-diffusion partial differential equations \cite{Clayton:2011}: 
\begin{align}
    \dd_t \uu = \Delta \bP \uu + \bF(\uu)+ \mathbf{h}. \label{RDE}
\end{align}
Here $\uu$ contains $N_{var}$ variables that depend on space and time, $\Delta \bP \uu$ selects which variables are spatially coupled by diffusion, $\bF(\uu)$ describes local excitability and recovery and the perturbation $\mathbf{h}$ can be used to capture external stimulation or interaction. The spiral wave patterns in Fig. \ref{fig:intro} show the spatial profile of one variable at given time $t_1$, e.g. $u_1(x,y,t_1)$. 

\section{Results}

\textbf{Spatial drift law for interacting spirals.} For a set of $N$ spiral waves in the plane (or a bounded part of it), we make use of their particle-like behavior and call the cartesian coordinates of their centers $\vec{X}_j = (X_j(t), Y_j(t))$, where $j=1,2,...,N$. The rotation phase $\Phi_j(t)$ is the angle over which a reference spiral needs to be rotated to best match the observed spiral wave at a given time $t$. 

\begin{figure}
    \centering
\includegraphics[width=0.5\textwidth]{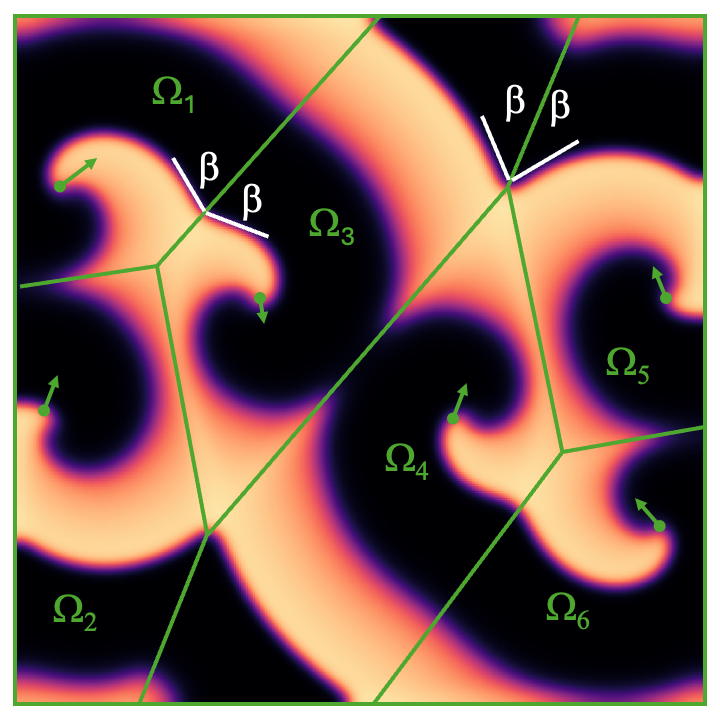}
    \caption{Unraveling complex excitation pattern by segmenting the pattern of Fig. \ref{fig:intro}c  into subdomains $\Omega_j$, each containing one spiral wave. The two waves meeting at the collision interfaces enclose the same angle $\beta$ at each side of the interface, which follows from the fact that their projected speeds along the interface must be equal. 
    \label{fig:subdomains}}
\end{figure}

We now spatially partition a complex non-linear pattern into its fundamental building blocks. Multiple sources of activation may be present including spiral waves and stimulation sites (e.g. pacemaker cells or external stimulation). The waves emanating from these sources naturally divide the domain in different subdomains, or regions of influence. Within each subdomain, the pattern resembles that of the corresponding source, but is spatially clipped to that region. Fig. \ref{fig:subdomains} shows such partitioning into subdomains, each governed by a single spiral wave.

The subdomains are separated by curves, which we will call collision
interfaces. Generally, the incoming wave fronts intersect the collision interface at an angle $\beta$ that varies along the interfaces, see examples drawn at Fig. \ref{fig:subdomains}. However, since the wave fronts meeting from both sides will travel together along the interface, angles $\beta$ at both sides must have the same value. Otherwise put, the collision interface bisects the wave front cusp where it crosses the interface. If $\beta = 90^\circ$, the wave front travels parallel to the boundary and the interaction is minimal. In contrast, if $\beta \approx 0^\circ$, the wave front intersects the boundary at a small angle, leading to a strong deflection of the wave front near the boundary. In the Methods section, we quantify these deflections and calculate their effect on the spiral cores. 

Our main result is the following. Let $\Omega_j$ denote the subdomain governed by the j-th spiral. Let $\vec{N}$ be the unit outward normal and $\vec{T}$ a counterclockwise unit tangent vector to the subdomain boundary $\partial \Omega_j$, parametrized by the arc length $s$. Then, the spiral's spatial drift is in leading order given by the sum of small forces generated at each point along the boundary, in the following way:
\begin{align}
    \mathbf{M}_j \cdot \frac{d\vec{X}_j}{dt} &= \oint_{\partial \Omega_j} [ F_N (r, \beta) \vec{N} + F_T(r, \beta)  \vec{T}] ds. \label{Newton-trans}
\end{align}

Here, $\mathbf{M}_j$ denotes the `mass' or `mobility' of the spiral, see below.  $F_N$ and $F_T$ are the normal and tangential components of the spiral wave interaction force. 

Equation \eqref{Newton-trans} parallels Newton's second law $M d^2\vec{X}/dt^2 = \vec{F}$ in classical mechanics. Since excitable media are dissipative, the underlying local equations \eqref{RDE} have first-order derivatives in time. Therefore, the forthcoming dynamical law is also first-order in time, implying that external influences (`forces') are proportional to the spiral wave drift velocity rather than its acceleration. Thus, spiral waves obey Aristotelian dynamics, which is a limit of Newtonian dynamics that is dominated by dissipative processes rather than inertia.

The two force components $F_N$ and $F_T$, can be understood from a rotation symmetry argument. A local disturbance at a point $P$ of on a segment of the subdomain boundary of length $ds$, will cause a drift of the spiral center $C$ relative to the line $CP$. In general, this net drift will not be directed along $CP$, similar to the drift response of a single spiral to parameter gradients. Decomposing this drift into components normal and tangential to the boundary yields the integrand in Eq. \eqref{Newton-trans}. Explicit expressions for the force densities $F_N$, $F_T$ developed along the collision interface are given in the Methods section. 

On the left-hand side of Eq. \eqref{Newton-trans}, a proportionality factor $M$ is found between cause ($\vec{F}$) and effect $(d\vec{X}/dt)$. This factor is similar to the mass in Newtonian mechanics, but differs in three ways. First, it couples to velocity, not acceleration, as explained above. Second, for finite domains, we find a mass tensor:
\begin{align}
    \mathbf{M}_j = \left(\begin{matrix}
        m_{||,j} & m_{\perp,j} \\
        -m_{\perp,j} & m_{||,j} \\
    \end{matrix} \right).
\end{align}
If $m_{perp}$ is non-zero, this means that the spiral wave reacts to a force exerted on it by moving in a direction that is not aligned with the force. (A somewhat similar effect can be seen when wind pushes a sailboat further.) In analogy to scalar and pseudoscalar filament tension \cite{Biktashev:1994}, this system this also possesses scalar mass $m_{||}$ and non-scalar mass $m_{\perp}$. Thirdly, the mass components depend on the region of influence:
\begin{align}
    m_{||,j} &= \iint_{\Omega_j} \rho_{||}(r_j) dS,&
    m_{\perp,j} &= \iint_{\Omega_j} \rho_{\perp} (r_j) dS.
\end{align}
In the limit of large domains, $m_{||} \rightarrow 1$, and $m_{\perp} \rightarrow 0$. However, for a small region of influence, both $m_{||}$ and $m_{\perp}$ will to zero, which could lead to faster drift response of spirals occupying a smaller surface area. 


\textbf{Rotational drift for interacting spirals.} Similarly to rigid body mechanics, spiral waves also possess a rotational degree of freedom. How their angular velocity changes from its value $\omega$ in the unbounded plane due to interactions at the collision interface is given by the equivalent of Euler's law $\tau = I d^2 \Phi / dt^2$: 
\begin{align}
    I_j \frac{d\Phi_j}{dt} &= I_j \omega + \oint_{\partial \Omega_j} \tau (r, \beta) ds.\label{Newton-rot}
\end{align}
where $I_j$ denotes the spiral's moment of inertia. As with the spiral scalar mass, $I_j$ decreases from 1 to 0 as the size of the subdomains $\Omega_j$ decreases. The torque $\tau$ is given by summing all contributions along the domain boundary, and again depends on the angle of intersection $\beta$ between the wave front and the collision interfaces surrounding its region of influence. 

\textbf{Solution to the spiral-boundary problem.} For a single spiral wave in a convex region $\Omega_1$ with no-flux boundary conditions, Eqs. \eqref{Newton-trans}-\eqref{Newton-rot} fully determine its dynamics, as the subdomain boundary is fixed. Our derivation assumes that, at all times, a continuous wave front connects the boundary to the spiral wave core region. In non-convex or elongated domains, this is not necessarily true, and further analysis is therefore required. Such regions may instead be treated as separate downstream subdomains without their own active sources. However, since spiral wave sensitivity decays exponentially with distance from its center, these regions tend to have only a small effect on spiral wave drift. 

Eqs. \eqref{Newton-trans}, \eqref{Newton-rot} also describe spiral wave behavior near an inexcitable obstacle or a hole in the domain. Here, the net balance of attraction vs. repulsion, together with the functional form of $F_N, F_T$ and $\tau$ determine whether spirals migrate to a hole or obstacle. This result generalizes the spiral interaction with localized defects, as established by Biktashev et al. \cite{Biktashev:2010}

\textbf{Location of collision interfaces.}
The case that motivated this study is spiral wave interaction, which requires determining the subdomains $\Omega_j$. To locate the collision interface between spiral $1$ and spiral $2$, we search for points that are reached simultaneously by their respective wave fronts. This can be achieved by constructing the isochrones of wave arrival time for the two spiral waves separately and identifying their intersections. This leads to a function
\begin{align}
    H(x,y,X_1,Y_1,X_2,Y_2) = \Phi_1 - \Phi_2. \label{subdomains}
\end{align}
In the Methods section, we show how to find $H$ from the wave front shape of a single unbound spiral in polar coordinates. The relative positions $(X_1, Y_1)$, $(X_2, Y_2)$ of the two spiral centers define a function $H(x,y)$ whose level curves show potential collision interface candidates. The specific interface selected depends on the relative phase difference between the spirals. Note that four distinct cases of $H$ arise, corresponding to the four possible combinations of spiral wave rotation senses. 

If several spirals are present with known positions and phase differences, one can solve Eq. \eqref{subdomains} for every pair, and define the spiral's region of influence as the set of points closer to that spiral than to any of the corresponding collision interfaces. This construction generalizes the concept of a Voronoi cell by incorporating the phase of each spiral into the definition of distance. 

\begin{figure}
    \centering
    \raisebox{2.5cm}{a)}
    \includegraphics[height=0.15\textwidth]{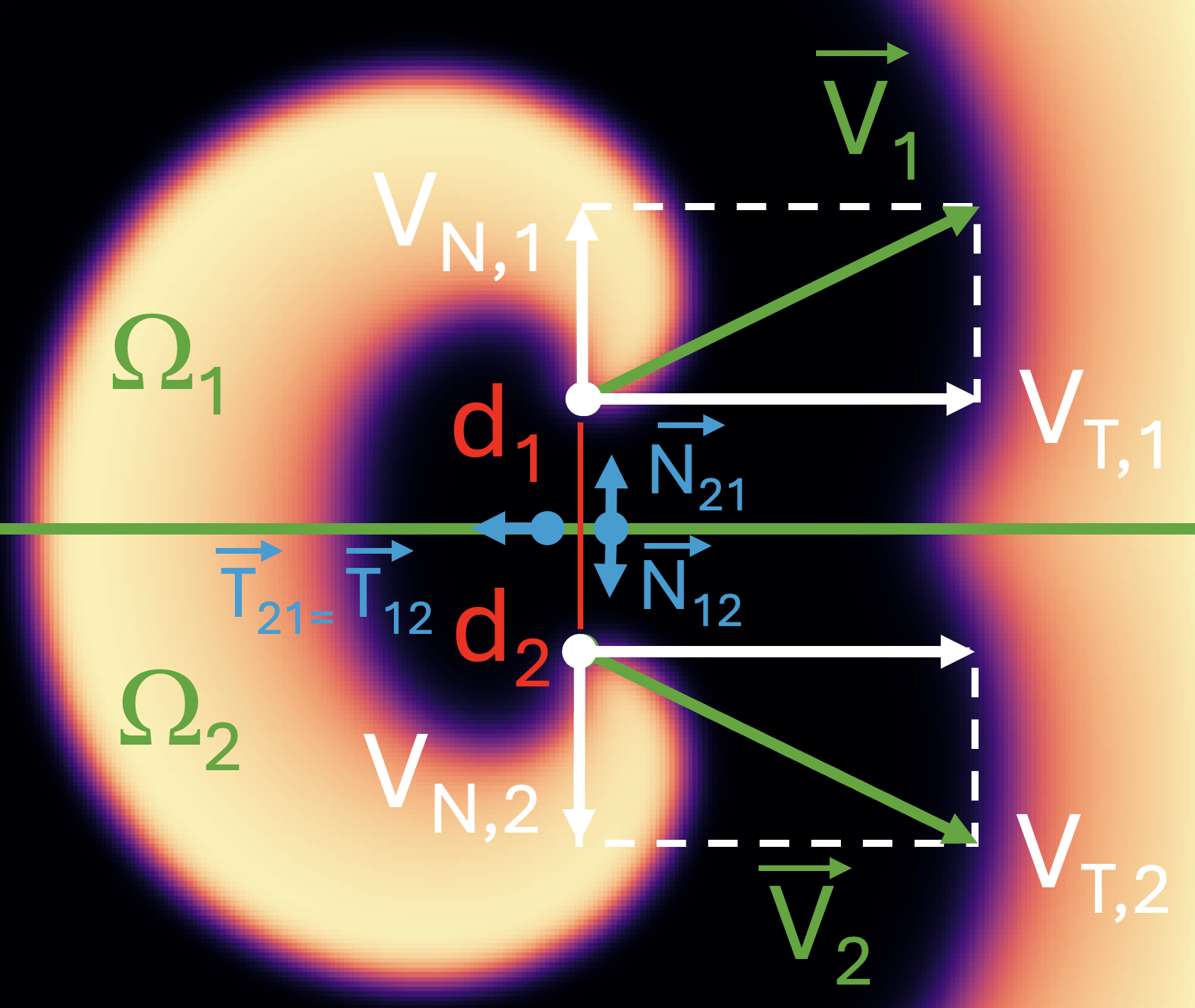}
  \raisebox{2.5cm}{b)}
    \includegraphics[height=0.15\textwidth]{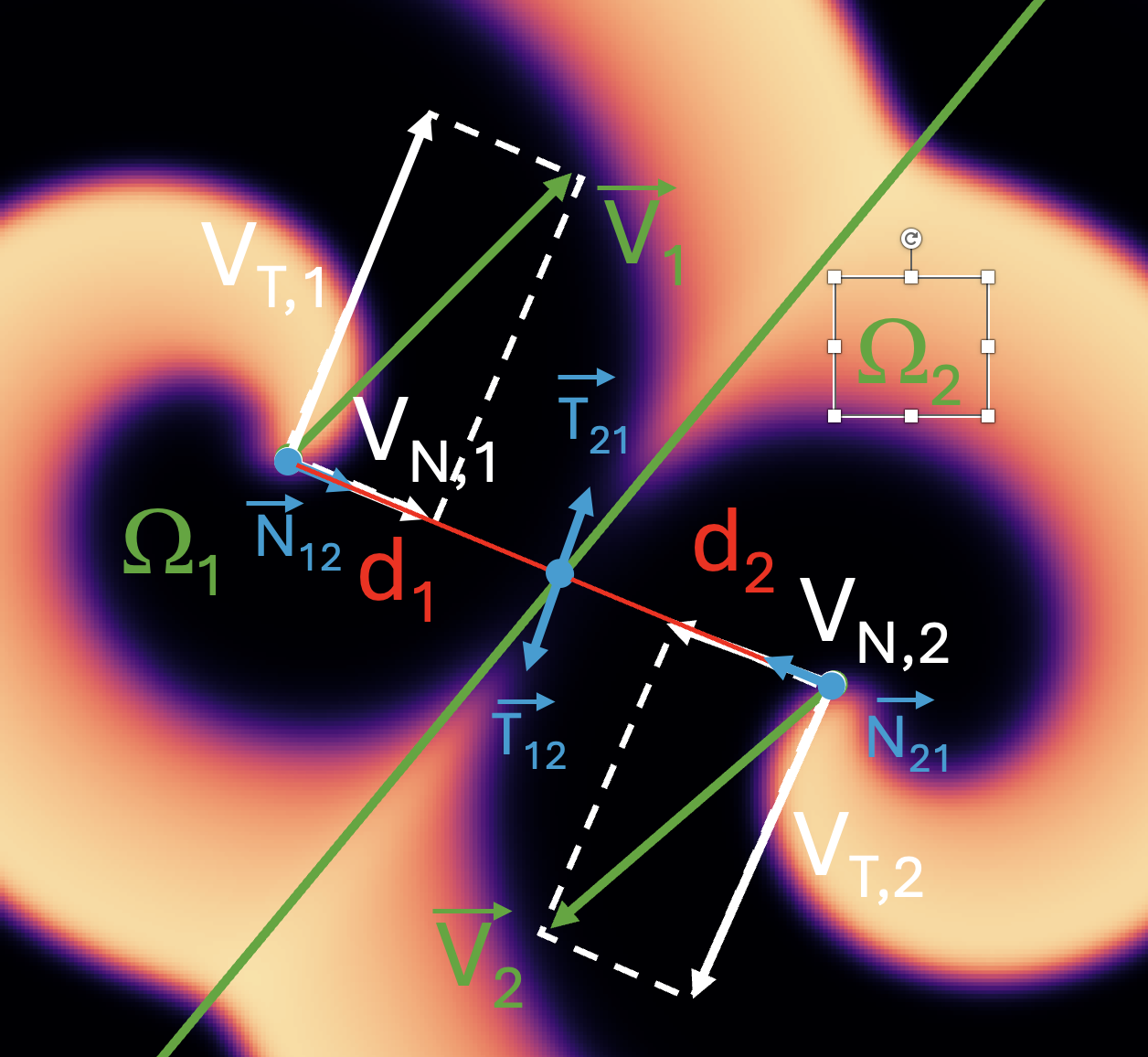}\\
  \raisebox{3cm}{c)}
    \includegraphics[width=0.2\textwidth]{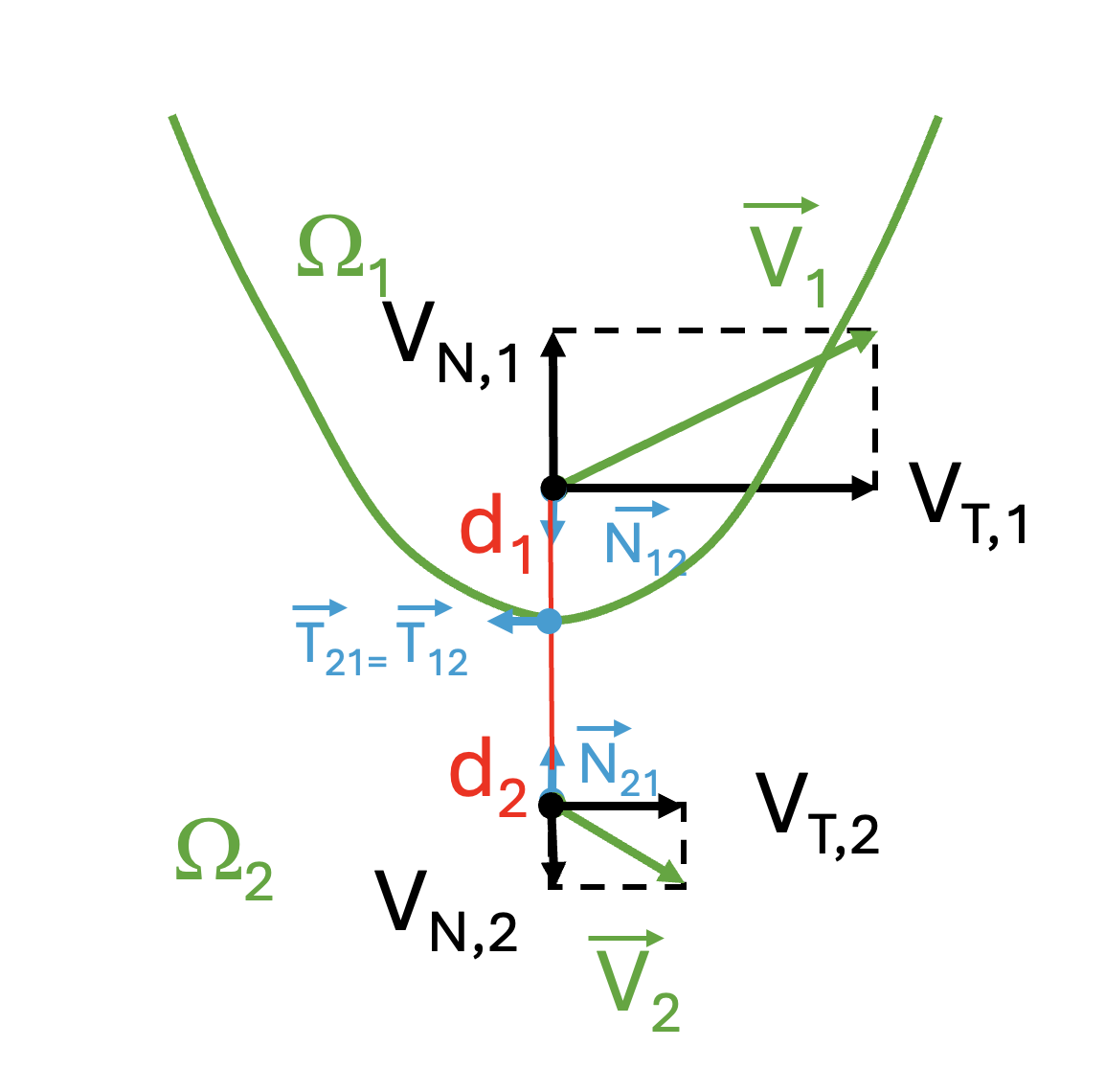}
  \raisebox{3cm}{d)}
    \includegraphics[width=0.2\textwidth]{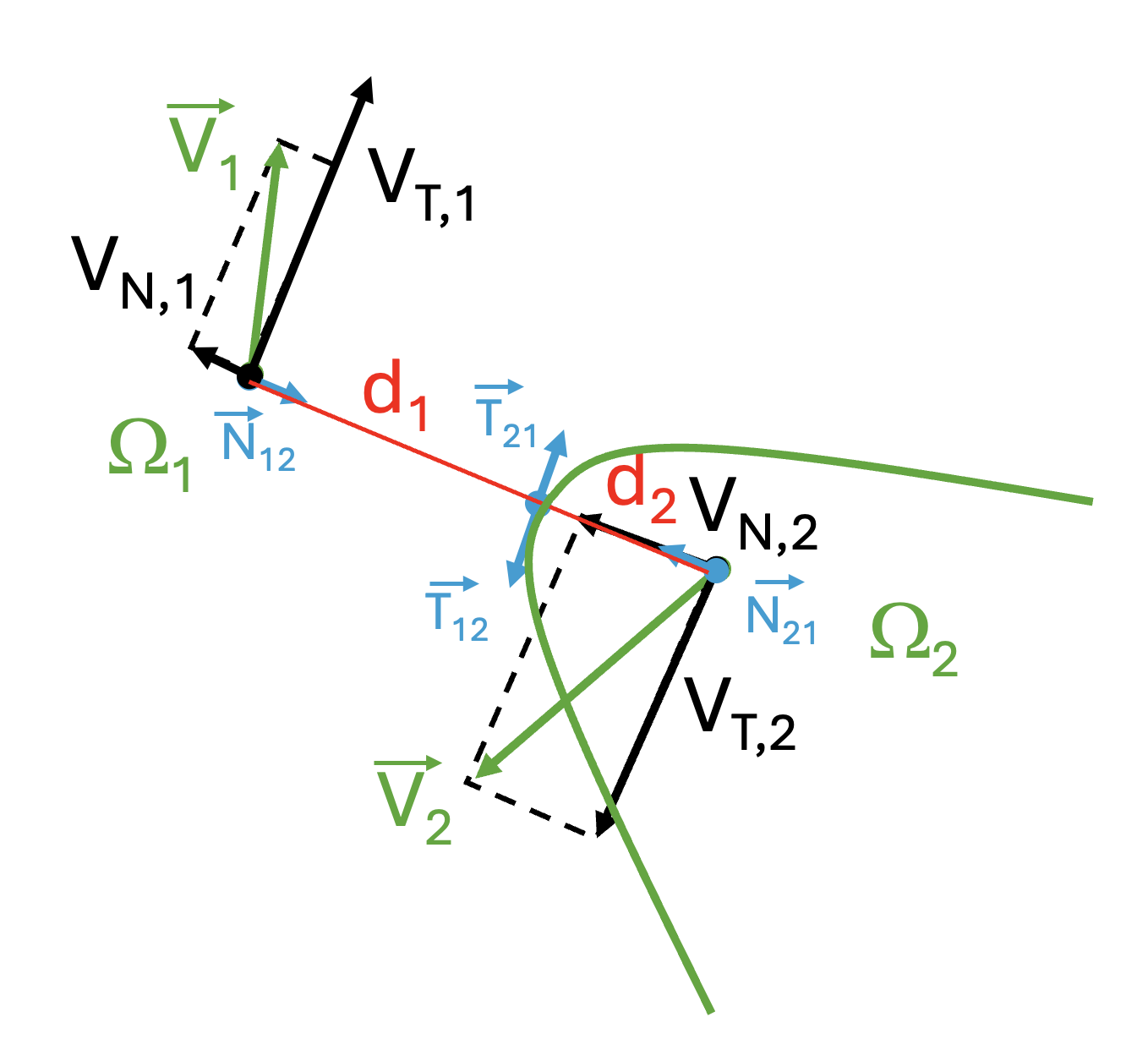}   
    \caption{Velocity components for the case of 2-spiral interaction: symmetric cases (a-b) vs. asymmetric cases (c-d). 
    (a) Oppositely rotating spirals with the same rotation phase.
    (b) Same-chirality spirals with opposite phases also create a straight collision interface. 
    (c) Oppositely rotating spirals with different rotation phase. The spiral lagging ahead occupies a larger region of influence.
    (d) Same-chirality spirals with different rotation phases. In cases (c) and (d), the reaction of one spiral to the other is not reciprocal, violating Newton's third law. 
    }
    \label{fig:twospirals}
\end{figure}

\textbf{Spiral pair dynamics.} As a first example, let us consider the case of two interacting spirals in an unbounded plane, see Fig. \ref{fig:intro}. Their dynamics will not only depend on the distance $r$ between the centers, but also on their respective distances $d_1$ and $d_2$ to the collision interface, see Fig. \ref{fig:twospirals}a-b. Then, the integrals \eqref{Newton-trans}, \eqref{Newton-rot} can be evaluated along the curves given by Eqs. \eqref{subdomains} yielding the drift components shown in Fig. \ref{fig:twospirals}:
\begin{subequations}
\begin{align}
    \frac{d \vec{X}_1}{dt} &= v_{N} (d_1, d_2) \vec{N}_{12} + v_T  (d_1, d_2) \vec{T}_{12}, \label{interaction_1}\\
     \frac{d \vec{X}_2}{dt} &= v_{N} (d_2, d_1) \vec{N}_{21} + v_T  (d_2, d_1) \vec{T}_{21},\label{interaction_2} \\
\frac{d\Phi_1}{dt} &= \omega +\tau(d_1, d_2),\\ 
\frac{d\Phi_2}{dt} &= \omega + \tau(d_2, d_1). 
\end{align}
\label{interaction}
\end{subequations}
Here, the factors $\mathbf{M}_j, I_j$ were brought to the right-hand side and absorbed in $v_N, v_T, \tau$.  

\textbf{A non-reciprocal force law.} The interaction laws \eqref{interaction} remain unchanged under exchange of labels $1$ and $2$, as expected from the invariance of physical laws under relabeling of two elements. Nonetheless, the net force of spiral 2 on spiral 1 (right-hand side of Eq. \eqref{interaction_1}) generally differs from the force exerted by spiral 1 on spiral 2 (Eq. \eqref{interaction_2}). Spiral waves therefore manifestly violate Newton’s third law, a feature that is characteristic of many non-equilibrium systems \cite{you_nonreciprocity_2020, dinelli_non-reciprocity_2023}. That the law of action and reaction is broken in spiral wave dynamics is already evident in simulations of pairs of oppositely rotating spirals: merely due to their interaction, the barycenter acquires a net total velocity $ \frac{d \vec{X}_1}{dt} + \frac{d \vec{X}_2}{dt}$. The origin of this violation lies in the fact that the force depends separately on $d_1$ and  $d_2$. By contrast, familiar conservative interactions such as gravitation and electrostatic interaction depend only on the total separation $d = d_1+d_2$, because they do not require an interface for force mediation. 

\textbf{Spiral pair dynamics.} In a co-moving frame of reference one can simplify spiral pair dynamics further to:
\begin{align}
    \frac{d (d_1)}{dt} &= f(d_1, d_2), &
    \frac{d (d_2)}{dt} &= f(d_2, d_1).
\end{align}
This symmetric 2 by 2 linear system allows a full stability analysis, revealing that (i) a symmetric spiral wave pair is generally stable, and (ii) asymmetric bound pairs can also exist (see \cite{zemlin_asymmetric_2005}), although their stability depends on conditions on the derivatives of $f$ at the equilibrium $d_1^*$ and $d_2^*$. 

\textbf{Generalization.} The presented results pertain to spiral wave interaction. In the presence of other external forces that do not create or destroy spirals, linear superposition can show the total response \cite{li_joint_2023}:
\begin{align}
    \mathbf{M}_j \cdot  \frac{d\vec{X_j}}{dt} &= \sum_k \vec{F}_{j,k}, &
    I_j \frac{d \Phi_j}{dt} &= \sum_k \tau_{j,k}.
\end{align}
Hence, although linear superposition is generally not applicable to non-linear dynamics, dividing the domain into subdomains with coherent patterns allows the total drift response to be calculated, which in turn will determine the collision interfaces. In the context of cardiac arrhythmias, tissue inhomogeneities can generate pinning forces and drift due to gradients in tissue properties, which may outweigh spiral wave interactions. In such cases, the process of spiral wave competition, as determined by Eq. \eqref{subdomains} may decide the long-term fate of the system, as discussed next.  

\textbf{Spiral competition.} A common rule-of-thumb in spiral wave scientific community (including cardiologists) is that the fastest source will eventually take over the system \cite{diagne_rhythms_2023}. This follows directly from Eq. \eqref{subdomains}: if $\Phi_1$ advances more rapidly than $\Phi_2$, the region of influence of $\Phi_1$  continually expands, providing a mathematical basis for the rule-of-thumb. Because the only medium-specific input in Eq. \eqref{subdomains} is the wave front shape, this competitive process is generic and does not depend on the details of excitability. Moreover, since spiral $1$ gains mass s its region of influence grows while spiral $2$ correspondingly loses mass, the dominance of the faster spiral is expected to take place at an accelerating pace. Due to the dependency of the spiral mass on its region of influence, its drift response can change significantly when its domain is almost vanishing, potentially preventing annihilation near the boundary in anti-tachycardia pacing. 

\begin{figure}
    \centering
    \includegraphics[width=0.5\textwidth]{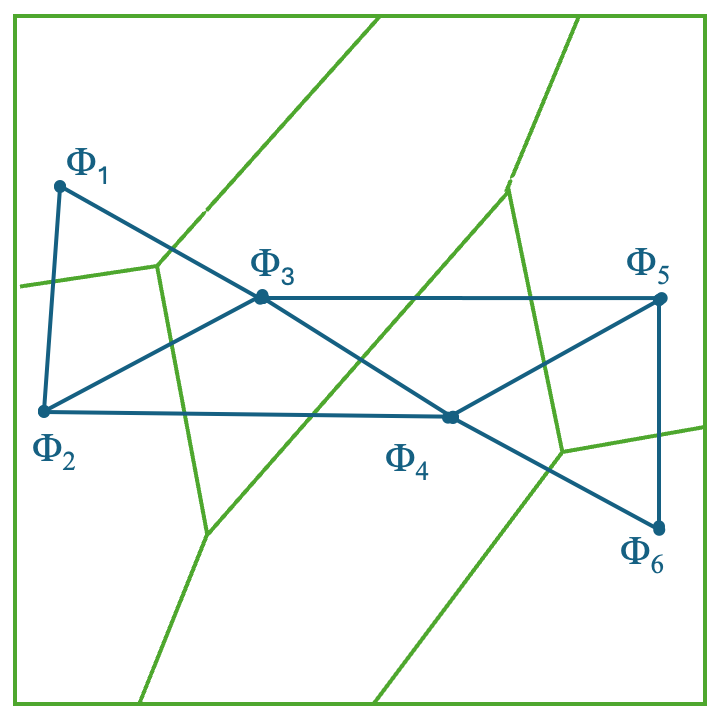}
    \caption{Representation of the multiple spiral state from Fig. \ref{fig:intro}c as a graph (blue). Every vertex represents a spiral center and a connecting edge is added if they interact directly, i.e. share a collision interface.}
    \label{fig:graph}
\end{figure}

\textbf{Transition between mother rotor fibrillation and multiple wavelet fibrillation.} The spiral interaction force is limited in magnitude by the extent of the spiral wave's sensitivity functions, which decay exponentially. However, phase differences tend to accumulate over time, and by Eq. \eqref{subdomains}, they can shift the collision interface even when the spirals are far apart. In a first, crude approximation, an arrangement of $N$ spirals on a surface can therefore be represented by a graph with $N$ nodes, where an edge connects two nodes whenever the corresponding spirals share a collision interface, see Fig. \ref{fig:graph}. This graph captures which spirals directly compete for territory and provides a simplified structure for analyzing how phase‑driven interface motion influences the system’s long‑term evolution.
Suppose these spirals are approximately regularly spaced and have phases $\Phi_j$ relative to their evolution without interaction. We assume that $\tau(d_1, d_2)$ depends more strongly on the first distance $d_1$, since $d_2$ primarily sets the curvature of the intersection boundary. Therefore, we take $\tau_j(d_j,d_k) / I_j \approx g(\Phi_j-\Phi_k)$. This leads to a simplified graph-based model of competing spirals during fibrillation:
\begin{align}
    \frac{d\Phi_j}{dt} = \omega + \sum_{k \in N_j} g(\Phi_j - \Phi_k). 
\end{align}
For $f = \sin$, this reduced to the well-known Kuramoto model for synchronization. 
Here, $N_j$ denotes the set of neighboring vertices of the j-the vertex. By symmetry, if the collision interfaces are always at the midpoint of the segments connecting spiral centers, the system is stationary by symmetry. Its stability depends on the eigenvalues of the Jacobian, which, in the case of equal coupling functions $g$ is simply the graph Laplacian matrix of the system. Since this matrix is positive semi-definite, the stability of this configuration depends on the sign of $g'(0)$, i.e. whether moving an interface away from a spiral center accelerates the spiral. If $g'(0) >0$, the equilibrium system is unstable, and eventually one spiral dominates the pattern. Conversely, if $g'(0) <0$, the configuration with collision interfaces at the midpoints of spiral centers is stable. These two scenarios correspond to the classical fibrillation regimes known in literature as mother-rotor fibrillation vs. multiple wavelet fibrillation \cite{Jalife:1998b,shibata_mechanism_2022}. While actual cardiac fibrillation involves many additional effects, we here propose that distinct disease states can be understood as emergent phase transitions in the language of physics, or equivalently, as bifurcations in the framework of dynamical systems theory. 

\section{Discussion}

In this work, we derive drift laws for spiral waves in general excitable and oscillatory systems. As an alternative to a purely numerical approach, our works bridges from the microscopic to the macroscopic scale via analysis of partial differential equations, relying on the spatial sensitivity of the spiral waves and taking integrals over the regions of influence, over spiral periods and along subdomain boundaries.
We show that spiral waves interact primarily with their nearest neighbors, with the interaction localized at the interfaces where their wave fronts collide. This collision boundary bisects the resulting cusps and moves over time, and the interaction angle together with the distance from the spiral core determine the resulting force. 

At the core of our result lies a mathematical derivation, detailed in the Methods section, in which an approximate solution to the multi-variable reaction-diffusion system is constructed by patching together single spiral solutions per subdomain. his analytical route—rather than a purely data‑driven one—makes it possible to identify the validity range of each intermediate approximation and to explain the observed phenomena in mechanistic terms. 

Several analogies with other physical systems can be noted. The particle-wave duality of spiral waves is used to find effective interaction laws. The region of space occupied by a spiral is reminiscent of magnetic domains and is separated by domain walls. In that sense, the transition of a single-spiral state (monomorphic tachycardia, mother rotor fibrillation) to a multi-spiral state (cardiac fibrillation) can be compared to the process of demagnetization of a material above the Curie temperature. 

This work opens several avenues for both fundamental and applied investigation. On the theoretical side, the asymptotic matching of non-linear patterns could be extended to develop a deeper fundamental geometric understanding of a a broader class of wave phenomena, including focal sources, meandering spirals, external stimulation and three-dimensional structures such as scroll waves. It also provides a framework for studying how spirals interact with holes, obstacles of intermediate size, and inhomogeneities across multiple spatial scales. 
On the applied side, the various dissipative forces predicted by the theory should be quantified experimentally. Measuring their relative magnitudes would help assess how accurately the present framework captures real systems and would clarify which mechanisms dominate under different physiological or physical conditions. Such measurements would also guide refinements of the model and help identify regimes where additional effects such as anisotropy, curvature or stochasticity become important.

In view of future medical applications, the geometric perspective reframes disease as a qualitative shift in system dynamics driven by nonlinear interactions across scales. By combining nonlinear modeling with rigorous mathematical analysis, it becomes possible to identify early‑warning signals, critical thresholds, and the mechanistic pathways that lead to pathological transitions. This provides a principled basis for predicting when a system is approaching instability, stratifying risk according to its proximity to a tipping point, and designing interventions that actively steer biological dynamics back toward stable regimes.
Integrating concepts from statistical physics and nonlinear dynamics into biomedical research promises to deepen our understanding of how diseases emerge and evolve. Such an approach supports the development of mechanism‑based therapeutic strategies that target the underlying dynamical landscape rather than only its downstream manifestations. As this perspective matures, it may help bridge the gap between mathematical theory and clinical decision‑making, offering new ways to interpret, anticipate, and ultimately modulate complex biological behavior.

\section{Conclusion}

We uncovered particle-like properties of spiral waves in non-linear systems: they possess an effective mass and moment of inertia determined by the extent of their region of influence. Their drift arises from Aristotelian forces, which in the context of spiral-spiral or spiral-boundary interactions originate from wave front deflections at the moving collision interfaces between neighboring spirals. Deriving physical laws of interaction opens new avenues to characterize tipping points in complex systems, here exemplified by the transition between different types of cardiac fibrillation. 

\section{Methods}

\subsection{Numerical methods.} 
Numerical simulations were performed using a finite-difference Euler timestepping of the reaction-diffusion equations with Aliev-Panfilov kinetics $\bF(\uu)$ \cite{Aliev:1996}, using the python software package FiniteWave (https://github.com/finitewave). 

\subsection{\textit{In vitro} experiments.} 

Experimental recordings in Fig. \ref{fig:intro} show the optical voltage mapping intensity \cite{Salama:1987} in cultured layers of conditionally immortalized human atrial myocytes, as developed by Harlaar et al. \cite{Harlaar:2021}. Visualization was performed using the Sappho python software module \cite{kabus_fast_2024}. 

\subsection{Mathematical derivation.} 

We construct an approximate multi-spiral solution to the reaction-diffusion equations in the following steps. 

(i) \textbf{Find the regions of influence}, given the positions $X_j, Y_j$ of the spiral wave centers and their phases $\Phi_j$. Let us call $(r_j, \theta_j)$ polar coordinates around the rotation center of the $j-$th spiral, and let $\theta_j = f(r_j)$ the shape of the clockwise rotating spiral's wave front in an unbounded domain without other spiral waves. 
In case of many spirals, we take the local activation phase $\phi_j(x,y,t)$ of the j-th spiral to be 
\begin{align}
  \phi_j(x,y,t) &=  \Phi_{j}(t) +  \omega t  - f(r_j) + \sigma_j \theta_j. 
\end{align} \label{phi12}
Here, $\theta_j = f(r_j)$ is the wave front shape of a single spiral solution to $\eqref{RDE}$ in polar coordinates. E.g. $\theta_j = k r_j$ would yield an Archimedean spiral. Furthermore, $\sigma_j=1$ for clockwise rotating spirals, and $-1$ for counterclockwise rotation. At the collision interface between the j-th and k-th spiral, activation phases must be equal (modulo $2\pi$). Setting $\phi_j = \phi_k$ locally, the collision interface turns out to satisfy Eq. \eqref{subdomains} with 
\begin{align}
H(x,y,X_j,Y_j, X_k, Y_k) = \sigma_j \theta_j + \sigma_k \theta_k + f(r_j) - f(r_k)
\end{align}
with $r_j = \sqrt{(x-X_j)^2+(y-Y_j)^2}$, $\tan \theta_j = (y-Y_j)/(x-X_j)$. The four possible choices of $\sigma_j = \pm 1$ depending on spiral wave chirality define different $H$. Only the cases of equal rotation and opposite rotation are fundamentally different, as the other 2 cases follow from a reflection around the line connecting the spiral centers.

\textbf{(iii) Set up the zeroth order multi-spiral solution.} We start from the unbound, unperturbed single spiral wave solution in the plane that rotates at angular frequency $\omega$. That is, $\UU(r,\phi(r,\theta,t)$ with $\phi$ given by Eq. \eqref{phi12}.

In polar coordinates we take $\UU(r,\phi + \varphi)$ with $\phi$ given by Eq. \eqref{phi12} and $\tilde{\phi}$ a phase shift that allows to match the single-spiral solutions at the collision interfaces. 

We here consider only the part of the phase shift that is constant during one spiral period, i.e. we assume that the spiral drift is slow, and that it fully accommodates its shape to the nearby spirals and boundaries. 

From substituting this form into Eqs. \eqref{RDE}, the phase shift is found to satisfy an advection-diffusion equation: 
    


\begin{align}
    0 &= \Gamma(r) \Delta \varphi + 2 A(r) \vec\nabla{\phi} \cdot \vec \nabla \varphi,
\end{align}
with boundary condition $\partial_n \varphi = - \partial_n \phi$. 

Since the spiral wave arms convey the perturbation outward, the solution takes a boundary layer structure near the collision interface. With $z$ the distance to the interface with arc length $s$, we find close to it that:
\begin{align}
 \varphi(s,z) = \frac{\delta(s) \omega}{c} \exp ( - \cos \beta(s) z / \delta(s)). 
\end{align}
with boundary layer thickness $\delta = \frac{c \Gamma}{2A}$. 

Finally, the shift in the spiral wave's coordinates follows from projecting onto the spiral wave's response functions $\WW^\nu$ \cite{Keener:1988,Barkley:1992,Henry:2000,Biktasheva:2009}: 
\begin{align}
\sum_\mu \frac{dX^\mu}{dt} \iint_{\Omega_j} (\WW^\nu)^H \partial_\mu \UU dS = \nonumber \\
\iint_{\Omega_j} (\WW^\nu)^H \dd_\Theta \UU \ \varphi(x,y) dS.  
\end{align}
Here, $\mu, \nu \in \{x,y,\Theta \}$. Due to the boundary layer structure, the right-hand side can be approximated by an integral over the collision interface. 
On the left hand side, one recovers after averaging over one spiral period the densities for the spiral's scalar mass, pseudoscalar mass and moment of inertia: 
\begin{align}
 \rho_{||} &=\frac{1}{4\pi} \oint d\theta (\WW^x \dd_x \UU + \WW^y \dd_y \UU), \nonumber\\
 \rho_{\perp} &=\frac{1}{4\pi} \oint d\theta (\WW^y \dd_x \UU - \WW^x \dd_y \UU), \\
 I_j &=\iint_{\Omega_j} d\theta \WW^\theta \dd_\theta \UU. \nonumber
\end{align}

\section*{References}
\bibliography{references_all}

\end{document}